\documentclass[twocolumn,prb,showpacs,preprintnumbers,amsmath,amssymb,floatfix]{revtex4}
\usepackage[dvips]{graphicx}
\usepackage{dcolumn}
\usepackage{bm}
\begin{document}
\newcommand{\Co}{La$_2$Co$_2$Se$_2$O$_3$}
\newcommand{\BaFe}{Ba$_2$Fe$_2$Se$_2$OF$_2$}
\newcommand{\LaFe}{La$_2$Fe$_2$Se$_2$O$_3$}

\title{Electronic structure, spin state, and magnetism of the square-lattice 
       Mott insulator La$_2$Co$_2$Se$_2$O$_3$ from first principles}

\author{Hua Wu}
  \affiliation{ II. Physikalisches Institut, Universit\"{a}t zu K\"{o}ln, 
               Z\"{u}lpicher Str. 77, 50937 K\"{o}ln, Germany}

\date{\today}

\begin{abstract}
{Electronic and magnetic structures of the newly synthesized cobalt oxyselenide
La$_2$Co$_2$Se$_2$O$_3$ (structurally similar to the superconducting iron pnictides)
are studied through density functional calculations.
The obtained results show that
this material is a Mott insulator, and that it has a very stable Co$^{2+}$ high-spin 
ground state with a $t_{2g}$-like orbital ordering, which is substantiated by
the calculated crystal-field excitation energies. 
The square lattice of the Co$^{2+}$ spins is found to have a strong antiferro (a weak 
ferro) magnetic coupling for the second nearest neighbors (2nn) via O (Se$_2$) and 
an intermediate 
antiferro one for the 1nn, with the strength ratio about 10:1:3.
The present results account for the available experimental data of magnetism, and 
the prediction of a planar frustrated (2$\times$2) antiferromagnetic structure 
would motivate a new experiment.}
  
\end{abstract}

\pacs{75.30.-m, 71.20.-b, 71.27.+a, 71.70.-d}

\maketitle


Very recently transition-metal oxychalcogenides receive revived 
interest~\cite{Mayer92,Kabbour08,Clarke08,Wang10,Zhu10} due to
their structural similarity to the superconducting iron 
pnictides~\cite{Kamihara08} in terms of a common square lattice of the
transition-metal species [with a local octahedral (tetrahedral)
coordination in the former (latter)].
In contrast to bad metal properties of the parent iron pnictides,
a Mott insulating behavior has been found in the iron or cobalt 
oxychalcogenides.~\cite{Kabbour08,Zhu10,Wang10} 
As the strength of electron correlations in the pnictides
is currently under a hot debate, and as the superconductivity and magnetism 
therein have a close relationship, it is worth a lot to study promptly the 
correlated-electron structure and magnetic properties of those Mott insulating
oxychalcogenides, which would help to understand the pnictides upon their
approaching the limit of a Mott localization. The structure and magnetic properties of 
(Ba,Sr)$_2$Fe$_2$(Se,S)$_2$OF$_2$ have recently been studied by Kabbour 
$et$ $al$.~\cite{Kabbour08} The observed long-range antiferromagnetic (AF)
ordering below 83.6-106.2 K was proposed to be a rare example of a frustrated
AF checkerboard spin lattice. More recently, Zhu $et$ $al$. studied 
a Mott localization in La$_2$Fe$_2$(Se,S)$_2$O$_3$ and found that
those materials have moderate charge gaps.~\cite{Zhu10} Their results support the notion
that the iron pnictides and chalcogenides possess intermediately strong
electron correlations and are not too far away from Mott localizations.

In this Rapid Communication, we study the cobalt oxyselenide {\Co}, 
which was newly synthesized by Wang $et$ $al$ and found by them
to be insulating and to have an AF transition at 
$T_{\rm AF}\sim$220 K.~\cite{Wang10}
Those authors suggested an unusual low-spin (LS, $S$=1/2) state of the 
Co$^{2+}$ ions and a corresponding orbital order for understanding 
of the insulating behavior and AF in the checkerboard spin-lattice. 
The Co$^{2+}$ LS state is quite surprising, in view of the common 
high-spin (HS, $S$=3/2) state.
Having a close look at the calculated results of Wang $et$ $al$. 
(Fig. 6 in Ref. \onlinecite{Wang10}) using the local-spin-density approximation
plus Hubbard $U$ (LSDA+$U$) method, one may wonder why three
singly occupied orthogonal orbitals have two up-spins and one
down-spin, which obviously violates the Hund's first rule. This infers
that their LSDA+$U$ calculation with an orbital-polarized potential ran into 
an exotic solution which should not be the ground state. Secondly, if the 
Co$^{2+}$ is in the LS state as Wang $et$ $al$ concluded using measurements of 
magnetic susceptibility and specific heat~\cite{Wang10} (which will be commented 
on below), the magnetic coupling between the $S$=1/2 ions, 
with a most probable frustration, would not be expected to yield the quite 
high $T_{\rm AF}\sim$220 K. Thus the electronic structure and magnetism of 
this interesting new material were poorly understood. For this reason, 
we carry out a rather complete set of density functional calculations
to study the electronic and magnetic structures of {\Co}, which
are closely related to the Co$^{2+}$ spin and orbital states.    
In strong contrast to the previous finding,~\cite{Wang10} 
all the present calculations using LSDA, LSDA-Fock-0.25, and LSDA+$U$ functionals 
consistently conclude the HS ground state, 
which is also substantiated by the calculated crystal-field level splittings. 
The orbital multiplet effect and a $t_{2g}$-like orbital ordering are discussed. 
Moreover, the exchange constants of the Co$^{2+}$ square lattice
are calculated and a planar frustrated (2$\times$2) AF structure is proposed. 
The present work
sheds new light on this interesting material, and a comparison
is also made with the above iron oxychalcogenides.

\begin{figure}[h]
 \centering\includegraphics[angle=0,width=8cm]{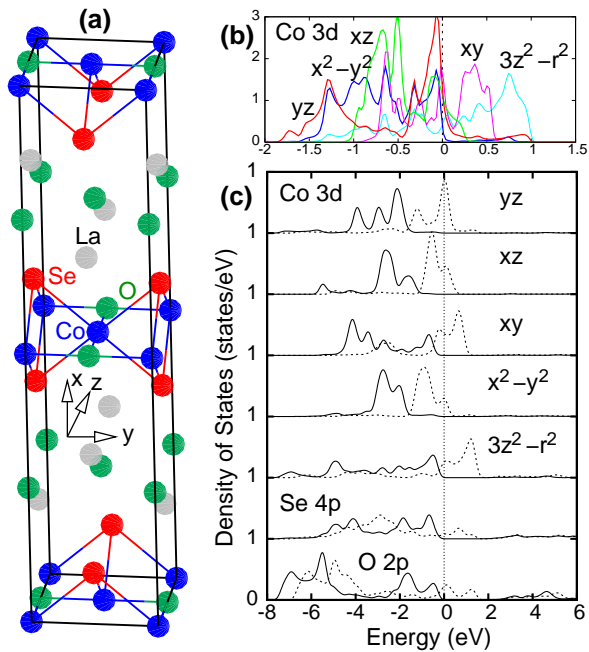}
 \caption{(Color online) (a) Crystal structure of {\Co} having the 
Co$_2$Se$_2$O layers. For each constituent CoSe$_4$O$_2$ octahedron the 
corresponding local 
coordinates are used, e.g., the $xyz$ for the body-centered Co site
and the $y$ and $z$ interchanged for the $c$-axis edge Co$^{2+}$. 
(b) Orbitally resolved Co 3d DOS for the nonmagnetic state by LDA.
The lower-lying $t_{2g}$-like orbitals ($x^2$--$y^2$, $yz$, and $xz$)
and the higher $e_g$ ($xy$ and $3z^2$--$r^2$) all split.
(c) DOS of the HS FM half-metallic state by LSDA. 
The solid (dashed) curves refer to the majority (minority) spin.
Fermi level is set at zero energy.}
 \label{fig1}
\end{figure}

{\Co} has the Co$_2$Se$_2$O layers which are separated by
the La$_2$O$_2$ layers, see Fig. 1(a).
Each Co$_2$Se$_2$O layer consists of the CoSe$_4$O$_2$ octahedra having
the Co-O (Co-Se) bondlength of 2.035 (2.688) \AA.~\cite{Wang10} Those
octahedra are
face-shared for the first-nearest-neighboring Co sites and edge- (corner-) 
shared via Se$_2$ (O) for the second-nearest Co neighbors. In our calculations,
we used for each CoSe$_4$O$_2$ octahedron such corresponding local $xyz$ coordinates 
that for the body-centered Co site as an example, the $z$ axis is along the 
Co-O bond and the $x$ along the crystallographic $c$ axis. As a result of 
the $D_{2h}$ point-group symmetry seen by the Co$^{2+}$ ions, 
the lower-lying $t_{2g}$-like orbitals ($x^2$--$y^2$, $yz$, and $xz$) and
the higher $e_g$ ($xy$ and $3z^2$--$r^2$) all split,~\cite{footnote1} see Fig. 1(b) 
and the detailed results below. 
Our calculations were carried out using the 
full-potential augmented plane waves plus local orbital method.~\cite{WIEN2k} 
The muffin-tin sphere radii are chosen to be 2.8, 2.2, 2.0 and 1.5 Bohr
for La, Co, Se, and O atoms, respectively.
The cut-off energy of 16 Ryd is used for plane wave expansion,
and 15$\times$15$\times$3 {\bf k} mesh for integrations over the Brillouin zone,
both of which ensure a high accuracy of our results.


We start with the standard LSDA calculations. The results show that independent of
the initialized spin moment, a ferromagnetic (FM) state always converges to the
same HS solution, see its density of states (DOS) in Fig. 1(c).
It is half metallic, with the fully occupied (and thus insulating) spin-majority
channel and partially occupied (metallic) Co $3d$ bands. This solution is calculated
to be more stable than the nonmagnetic state by 593 meV/f.u. We also did for the
assumed LS state a fixed-spin-moment calculation (otherwise it converges to the above
HS ground state) and find that the LS-fixed state is less stable than the HS state
by 435 meV/f.u., see Table I. Moreover, our calculations show that the HS FM state
is more stable than the HS checkerboard-AF state by 26 meV/f.u. This is most probably
due to a maximal kinetic energy gain in the FM half-metallic state. Note, however, 
that this FM half metallic ground state disagrees with the
experimental insulating and AF state. By a comparison with LaFeAsO 
(La$_2$Fe$_2$Se$_2$O$_3$) in which the Fe $3d$ bandwidth is 
4.4 (3.2) eV,~\cite{Zhu10}
the entire Co $3d$ bandwidth in {\Co} is only 2.8 eV, see Fig. 1(b).
This signals a stronger correlation effect in {\Co} due to the
band narrowing.~\cite{Zhu10} Indeed a presence of the local magnetic moment 
suggests a Mott localization behavior of {\Co}.~\cite{Wang10} Therefore, it is not
surprising that the standard LSDA calculations (albeit quite informative) 
cannot provide a correct description of the 
electronic and magnetic structure of this Mott insulator.         

\begin{table}
\caption{Relative total energies in unit of meV/f.u. (and the Co spin moment in unit
of $\mu_B$) of {\Co}  
having the HS or LS Co$^{2+}$ ions in the checkerboard AF or FM state.
The three functionals all show the HS ground state. The symbol ``$\rightarrow$" 
stands for ``converges to".}
 \label{TableI}
\begin{tabular} {l@{\hskip0.5cm}c@{\hskip0.3cm}c@{\hskip0.4cm}c} \\ \hline\hline
 & LSDA & LSDA-Fock-0.25 & LSDA+$U$(4 eV) \\ \hline
HS-AF & 26 ($\pm$1.97) & 0 ($\pm$2.63) & 0 ($\pm$2.55) \\ 
HS-FM & 0 (2.38) & 258 (2.65) & 55 (2.58) \\  
LS-AF & $\rightarrow$HS-AF & 1503 ($\pm$0.91) & $\rightarrow$HS-AF \\
LS-FM & 435 (0.93) & 1527 (0.92) & $\rightarrow$HS-FM \\
\hline\hline
\end{tabular}
\end{table}

In order to study the electron correlation effect, we carried out hybrid functional
calculations using the LSDA-Fock-0.25 which mixes 25\% Fock exact exchange into the
LSDA exchange functional and is applied to the Co $3d$ electrons. 
This was demonstrated to be an alternative way (to the
LSDA+$U$, see below) for studying correlated electron 
systems.~\cite{Novak06,Tran06}
Our hybrid functional calculations show again that the Co$^{2+}$ HS is the ground
state, being much more stable than the LS state by about 1.5 eV/f.u., see Table I. 
[The corresponding much smaller value of about 0.4 eV/f.u. given by the LSDA may 
well be due to
the artificial (half-) metallic band structure which significantly suppresses 
the
energy splitting of the spin and orbital multiplets of the Co$^{2+}$ 
ions.~\cite{Wu10}]
Fig. 2 shows the orbitally resolved DOS for the HS (left panel) and LS (right panel)
solutions in the checkerboard AF state. The HS Mott insulating solution has 
one $t_{2g}$ hole on the minority-spin $xz$ orbital and two
$e_g$ ($xy$ and $3z^2$--$r^2$) holes, thus forming a $t_{2g}$ orbital ordering. 
This is the ground-state solution of the
orbital multiplets for the HS Co$^{2+}$ ions, as confirmed by the following 
LSDA+$U$ calculations.
In this HS insulating state, the AF exchange and superexchange are expected to be
dominant, and indeed our calculations show that the HS checkerboard
AF state is more stable than the FM state by 258 meV/f.u., see Table I.
In contrast, the very unstable LS state has the $t_{2g}^6e_g^1$ configuration
with the $e_g^1$ being an $xy$ electron, fully according with the crystal-field
level sequence calculated below. 
The LS AF state is (a little) more stable than the LS FM state,
but the energy difference of 24 meV/f.u. is less than one tenth of that 
for the HS state. Although exchange constants may well be overestimated by a 
hybrid functional,~\cite{Moreira02} those two values confirm, as expected, 
that the LS state would yield much weaker magnetic couplings than the HS state.
       
\begin{figure}
 \centering\includegraphics[angle=0,width=8cm]{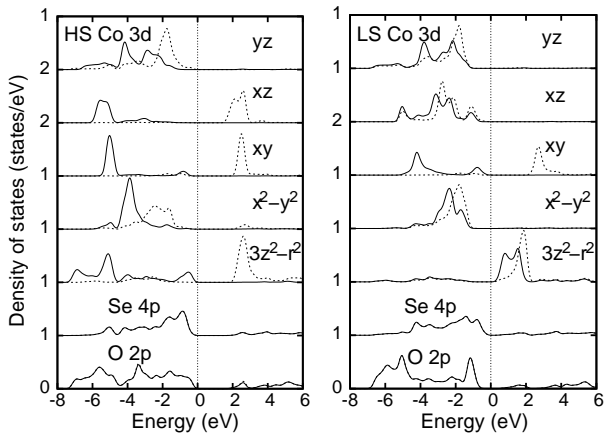}
\caption{DOS of the HS (left panel) and LS (right panel) solutions in the checkerboard
AF state by the LSDA-Fock-0.25 hybrid functional. The Co$^{2+}$ ions in the HS ground 
state have an $xz$-type ($t_{2g}$ hole) orbital ordering. 
The much less stable LS state has the $t_{2g}^6$+$e_g^1$(:$xy$) 
configuration. See
Fig. 1 for other notations.}
 \label{fig2}
\end{figure}

As the LSDA+$U$ method~\cite{Anisimov93} has been widely used to 
study the electronic and magnetic properties of the correlated electron systems, 
here we further performed a set of LSDA+$U$ calculations using
the double-counting correction in an atomic limit and the effective 
$U$=4 eV.~\cite{footnote2} 
We have initialized the corresponding density matrix (and orbital-dependent potential)
and then done self-consistently a full electronic relaxation. 
This procedure turns out to be suitable for study of the spin and orbital 
physics.\cite{Wu10}
The obtained electronic structure, the HS ground state, the orbital multiplet 
splitting, the exchange constants and the ground-state magnetic structure are 
discussed as follows. 

For both the FM and checkerboard AF configurations, the LS state turns out to be unstable 
and converges to the HS ground state. For the HS ground state, we tested all the ten 
different orbital configurations of the two minority-spin electrons (the majority-spin
channel is closed in the Co$^{2+}$ HS ground state). We find that the five lowest-energy
configurations states, the ($x^2$--$y^2$)($yz$), 
($x^2$--$y^2$)($xz$), ($yz$)($xz$), ($yz$)($xy$), and ($x^2$--$y^2$)($3z^2$--$r^2$),
order as 0, 60, 457, 750, and 790 meV/Co
(see Fig. 1 for orbital indices).  
Thus, the ground state of the orbital multiplets is 
$t_{2g\uparrow}^3$$e_{g\uparrow}^2$($x^2$--$y^2$)$_{\downarrow}^1$($yz$)$_{\downarrow}^1$, confirming the above hybrid functional results.
The corresponding DOS results (not shown here) are very similar to the latter. 
As each Co$^{2+}$ ion has a $t_{2g}$ hole on the $xz$ orbital in its local
coordinates and the local coordinates are rotated around the crystallographic
$c$-axis by $\pi$/2, to say, from the body-centered Co$^{2+}$ site to
the $c$-axis edge one [see Fig. 1(a)], each Co$_2$Se$_2$O layer has an 
antiferro $xz$-type
hole-orbital ordering. Moreover, from those above energy differences and 
by neglecting the difference of the inter-orbital Coulomb interactions,
the crystal-field level splittings (or to say, crystal-field 
excitation energies) 
are estimated: the $x^2$--$y^2$, $yz$, $xz$, $xy$, and  $3z^2$--$r^2$
order as 0, 397, 457, 750, and 1187 meV. By a comparison with the LS state ($t_{2g}^6$($xy$)$^1$),
the HS ground state ends with an $xz\rightarrow$$3z^2$--$r^2$ transition which gains, by a simple
estimate, two times Hund exchange (in total about 2 eV) with a much lower cost of the crystal-field
energy of 730 meV. This is the reason that the HS is the very stable ground state of this
Mott insulator, having a localized spin moment larger than 2.5 $\mu_B$ as seen 
in Table I.~\cite{footnote3}

Now we estimate the three exchange constants of the HS Co$^{2+}$ ions in the square lattice (Fig. 3),
by calculating four different magnetic structures (the same as used in Ref. \onlinecite{Kabbour08}, see Fig. 9 therein)
and then mapping the total-energy differences (i.e., exchange energies) 
onto a Heisenberg model.
The resultant values are --9.97 meV ($J_1$, AF, counted once per pair), --3.06 meV ($J_2$), and
0.98 meV ($J_3$, FM) for the second-nearest-neighboring (2nn) Co pair via O, the first-nearest-neighboring (1nn) pair,
and the 2nn pair via Se$_2$, respectively, as shown in Fig. 3. 
As the $J_1$ is more than three times as big as the $J_2$ (well meeting the requirement
$\left|J_1\right|>\left|J_2\right|$) and the $J_3$ is positive, the ground state 
AF structure 
[$E_{ex}$ = ($J_1$--$J_3$)$S^2$ per Co$^{2+}$ with $S$=3/2] is 
what we show in Fig. 3: It is more stable (albeit the frustrated $J_2$ couplings) 
than the checkerboard 
AF state [$E_{ex}$ =  (--$J_1$+2$J_2$--$J_3$)$S^2$ per Co] by 31 meV/Co, and than 
the stripe AF state [$E_{ex}$ = ($J_1$+$J_3$)$S^2$] by 4 meV/Co.
We also calculated those $J$ values using a smaller $U$ (=3 or 2 eV) when the system approaches
an itinerant limit, and find that while the FM $J_3$ becomes a bit smaller, both the AF $J_1$ 
and $J_2$ 
get larger with a nearly constant $J_1$/$J_2$ ratio of about 3. Thus, the 
HS planar (2$\times$2) AF structure as shown in Fig. 3 remains to be the ground state 
as the system is a Mott insulator with the effective $U\ge$ 2 eV. This prediction
calls for an experimental study.   

\begin{figure}
 \centering\includegraphics[angle=0,width=7.5cm]{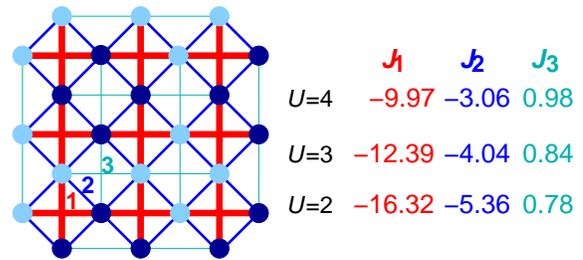}
\caption{(Color online) The ground-state (2$\times$2) planar magnetic structure
of the HS Co$^{2+}$ square lattice, with the 2nn AF exchange $J_1$ via 
O (red bold lines), the 1nn AF $J_2$ (blue solid lines), and the 2nn 
FM exchange $J_3$ via Se$_2$ (cyan thin lines). See also Fig. 1(a).
The $J_2$ couplings are
frustrated. Only the spin-up and -down Co$^{2+}$ sites are shown
for clarity. The listed exchange constants (in unit of meV) 
are calculated using the LSDA+$U$ with effective $U$=4, 3, or 2 eV.} 
 \label{fig3}
\end{figure}

Here we compare our LSDA+$U$ results with those calculated previously for the isostructural {\BaFe} and {\LaFe}.~\cite{Kabbour08,Zhu10} All the three materials are calculated to be a Mott insulator with a band 
gap opening by Hubbard $U$, and in the Mott insulating phase each of the three exchange constants
has the same sign (AF or FM) in the three materials. 
Both the Fe$^{2+}$ ($S_{\rm Fe}$=2) and Co$^{2+}$ ($S_{\rm Co}$=3/2) ions are found to be in the HS state.
Considering the strongest AF coupling for the Co or Fe pair via O, $J_1S_{\rm Co}^2$ in {\Co} 
is about 1.5 times as big as 
$J_1S_{\rm Fe}^2$ in {\BaFe} (Ref. \onlinecite{Kabbour08}) and 
$J'_2S_{\rm Fe}^2$ in {\LaFe} (Ref. \onlinecite{Zhu10}) (using those $J$ values 
at the $U$=3 eV for a direct comparison).
Moreover, the magnetic frustration appears weaker in {\Co} due to the smaller ratio $J_2$/$J_1<$ 1/3
(the corresponding ratio $>$0.5 in {\BaFe} and {\LaFe}.~\cite{Kabbour08,Zhu10})
Both factors account for the observations that the $T_{\rm AF}\sim$220 K 
in {\Co}
(Ref. \onlinecite{Wang10}) is about two times as high as that of around 100 K in both {\BaFe} and {\LaFe}.~\cite{Kabbour08,Zhu10} Note that if 
the Co$^{2+}$ was in the LS state, the $T_{\rm AF}$
would decrease probably by one order of magnitude as discussed above 
by the hybrid functional calculations. This would obviously disagree with the experiments.
Therefore the LS-state model is discarded by the present work.   

Now we briefly comment on the recent experiments carried out by Wang $et$ $al$.
for {\Co}.~\cite{Wang10} They used a Curie-Weiss law to fit the magnetic susceptibility in the quite 
narrow range of 317 K $<T<$ 400 K, and they also estimated a change of the magnetic entropy 
below 255 K, both of which led them to a conclusion of the unusual LS state.
As this layered material has a two-dimensional spin square-lattice with a considerably
strong magnetic frustration as found in the above calculations, the observed 
$T_{\rm AF}\sim$ 220 K may well be lower than the dominating AF coupling strength 
which is suppressed by other
frustrated magnetic interactions. This brings about a broad maximum of the susceptibility above
the $T_{\rm AF}$ as observed, and thus the above fitting using a Curie-Weiss law 
is less reliable:
the strongest AF coupling is not yet well broken, leading to an underestimate of the effective
moment. For the same reason, the decreasing magnetic entropy below 255 K represents only a partial but not
a full contribution of the spin degree of freedom, leading again to an underestimate of the spin degeneracy. 
As the HS ground state is very stable as we find above, and it has no room for a transition 
to a higher spin state with increasing temperature, a magnetic susceptibility 
measurement at a higher temperature, e.g., up to about 600 K, 
would be able to check it.      


To conclude, we studied the electronic structure, spin state, and magnetism
of the newly synthesized square-lattice cobalt oxyselenide {\Co}, using the
LSDA, LSDA-Fock-0.25, and LSDA+$U$ calculations. 
While the LSDA predicts this material to be a possible 
ferromagnetic half metal, both the LSDA-Fock-0.25 hybrid functional and 
LSDA+$U$ calculations show a Mott insulating and antiferromagnetic ground state 
in agreement with experiments, and a $t_{2g}$ orbital ordering as well. 
The exchange constants of the frustrated antiferromagnetism are calculated. 
In contrast to the newly suggested unusual 
low-spin state of the Co$^{2+}$ ions, all the present calculations 
conclude the very stable high-spin ground state, which is substantiated by the calculated
crystal-field level splittings. A comparison with the iron oxychalcogenides
is made, and the experiments are explained.
 
The author thanks Thomas Lorenz for helpful discussion.
This work is supported by the Deutsche Forschungsgemeinschaft
through SFB 608.


\begin{thebibliography}{50}

\bibitem{Mayer92} J. M. Mayer, L. F. Schneemeyer, T. Siegrist, J. V. Waszczak,
and B. Van Dover, Angew. Chem. Int. Ed. Engl. {\bf 31}, 1645 (1992).

\bibitem{Kabbour08} H. Kabbour, E. Janod, B. Corraze, M. Danot, C. Lee, 
M. H. Whangbo, and L. Cario, J. Am. Chem. Soc. {\bf 130}, 8261 (2008). 

\bibitem{Clarke08} S. J. Clarke, P. Adamson, S. J. C. Herkelrath, O. J. Rutt,
D. R. Parker, M. J. Pitcher, and C. F. Smura,
Inorg. Chem. {\bf 47}, 8473 (2008).

\bibitem{Wang10} C. Wang, M. Q. Tan, C. M. Feng, Z. F. Ma, S. Jiang,
Z. A. Xu, G. H. Cao, K. Matsubayashi, and Y. Uwatoko,
J. Am. Chem. Soc. {\bf 132}, 7069 (2010). 

\bibitem{Zhu10} J. X. Zhu, R. Yu, H. Wang, L. L. Zhao, M. D. Jones,
J. Dai, E. Abrahams, E. Morosan, M. Fang, and Q. Si, 
Phys. Rev. Lett. {\bf 104}, 216405 (2010). 

\bibitem{Kamihara08} Y. Kamihara, T. Watanabe, M. Hirano,
and H. Hosono, J. Am. Chem. Soc. {\bf 130}, 3296 (2008). 

\bibitem{footnote1} This eigenorbital set corresponds to
$x'y'$, $x'z$+$y'z$, $x'z$--$y'z$, $x'^2$--$y'^2$, and $3z^2$--$r^2$, 
when the $xyz$ is rotated around the $z$ by $\pi$/4 to the $x'y'z$.
Note that due to the lattice symmetry, the $x'z$ and $y'z$ are not eigenorbitals. 

\bibitem{WIEN2k} P. Blaha, K. Schwarz, G. Madsen, D. Kvasnicka, and J. Luitz,
Wien2k package, http://www.wien2k.at

\bibitem{Novak06} P. Nov\'{a}k, J. Kune\v{s}, L. Chaput, and W. E. Pickett, 
phys. stat. sol. (b) {\bf 243}, 563 (2006). 

\bibitem{Tran06} F. Tran, P. Blaha, K. Schwarz, and P. Nov\'{a}k, 
Phys. Rev. B {\bf 74}, 155108 (2006).

\bibitem{Wu10} H. Wu, Phys. Rev. B {\bf 81}, 115127 (2010).

\bibitem{Moreira02} I. de P. R. Moreira, F. Illas, and R. L. Martin,
Phys. Rev. B {\bf 65}, 155102 (2002).

\bibitem{Anisimov93} V. I. Anisimov, I. V. Solovyev, M. A. Korotin, 
M. T. Czy\.{z}yk, and G. A. Sawatzky,
Phys. Rev. B \textbf{48}, 16929 (1993).

\bibitem{footnote2} This corresponds to the value
of Hubbard $U$=5 eV subtracted by the Hund exchange of 1 eV. 
We find that the effective $U$=2 eV just opens a zero band gap, 
and that our conclusions remain unchanged upon the tested effective $U$=2-4 eV.

\bibitem{footnote3} In strong contrast, the metallic iron pnictides have an itinerant 
weak magnetism, the origin of which remains to be an open question [see $e.$$g.$,
I. I. Mazin and M. D. Johannes, Nature Phys. {\bf 5}, 141 (2009); F. Cricchio,
O. Gr{\aa}n\"{a}s, and L. Nordstr\"{o}m, Phys. Rev. B {\bf 81}, 140403(R) (2010);
E. Bascones, M. J. Calder\'{o}n, and B. Valenzuela, Phys. Rev. Lett. {\bf 104},
227201 (2010)].  

\end{thebibliography}
\end{document}